\documentclass[11pt]{article}
\pdfoutput=1

\usepackage{jheppub}
\usepackage{graphicx}
\usepackage{amssymb}
\usepackage{amsmath,amssymb}
\usepackage{slashed}
\usepackage{hyperref}
\usepackage{caption}
\usepackage{xcolor}
\usepackage{dsfont}
\usepackage{verbatim}
\usepackage{subfig}
\usepackage{fnpos}
\usepackage{ftnxtra}

\usepackage{tikz}
\usetikzlibrary{arrows, arrows.meta}

\newcommand\beq{\begin{equation}}
\newcommand\eeq{\end{equation}}
\newcommand\be{\begin{equation}}
\newcommand\ee{\end{equation}}

\title{Universal relations for Holographic Interfaces}

\preprint{\today}

\author[a]{Andreas Karch}
\author[b]{, Zhu-Xi Luo}
\author[a]{, Hao-Yu Sun}

\affiliation[a]{University of Texas, Austin, Physics Department, Austin TX 78712, USA}
\affiliation[b]{Kavli Institute for Theoretical Physics, University of California, Santa Barbara, CA, 93106, USA}

\emailAdd{karcha@utexas.edu,zhuxi\_luo@kitp.ucsb.edu,hkdavidsun@utexas.edu}

\abstract{We study the entanglement entropy in 1+1 dimensional conformal field theories in the presence of interfaces from a holographic perspective. Compared with the well-known case of boundary conformal field theories, interfaces allow for several interesting new observables. Depending on how the interface is located within the entangling region, the entanglement entropies differ and exhibit surprising new patterns and universal relations. While our analysis is performed within the framework of holography, we expect our results to hold more generally. 
}

\begin{document}
\maketitle

\section{Introduction and Summary}
\label{sec:intro}

Starting with the work of Cardy \cite{Cardy:1986gw}, 1+1 dimensional conformal field theories (CFTs) in the presence of boundaries have found many applications within the theory of critical phenomena as well as string theory. Somewhat less explored is the theory of conformal interfaces \cite{Bachas:2001vj}. An interface is a setup where two conformal field theories, each defined on a half line, meet at a pointlike defect across which they can communicate. The two conformal field theories on the two sides of the interface are often taken to be the same , i.e., the interface is simply a defect. But this needs not be the case. The more general setup involves different CFTs on both sides, and is similar to, say, the classic textbook case of 3d electrostatics of two dielectrics separated by a planar interface.

Interface conformal field theories (ICFTs) preserve the same symmetries as a boundary conformal field theories (BCFT) do. In fact, ICFTs can in principle be reduced to BCFTs via the so-called ``folding trick": consider one CFT living on the $x>0$ half space and the other on $x<0$. We can map the theory living at $x<0$ into the $x>0$ half by simply performing the operation $x \rightarrow -x$ on the left hand side of space; we folded space into a single half-line with a boundary. In this way we generated a BCFT albeit with a very special structure: in the bulk of the BCFT the Lagrangian describes two completely decoupled CFTs. The two only couple at the boundary. 
\begin{figure}[htbp]
\centering
\begin{tikzpicture}
    \draw[thick,blue] (2.5,0)--(2.5,3);
    \draw (0,0)--(5,0)--(5,3)--(0,3)--(0,0);
    \node at (1.2,1.5) {\small CFT$_L$};
    \node at (3.7,1.5) {\small CFT$_R$};
    \draw[<->] (5.5,1.5) -- (6,1.5);
    \draw (6.5,0)--(9,0)--(9,3)--(6.5,3);
    \draw[thick,blue] (6.5,0)--(6.5,3);
    \node at (7.75,1.5) {\small CFT$_L\otimes\overline{\text{CFT}}_R$};
\end{tikzpicture}
\caption{The folding trick. CFT$_L$ and CFT$_R$ connected through the blue conformal interface can be viewed as a conformal boundary theory (BCFT) for  the product CFT$_L\otimes\overline{\text{CFT}}_R$, where $\overline{\text{CFT}}_R$ means that left- and right-movers of CFT$_R$ are exchanged.}
\label{fig:fold}
\end{figure}
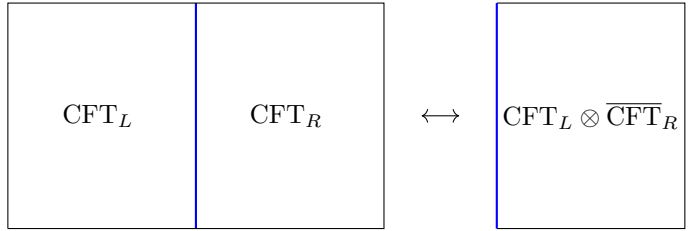
A simple, illustrative example of a 1+1 dimensional ICFT is the Ising defect CFT, an Ising model where the nearest neighbor coupling between the spins takes a slightly different strength on one link -- across the interface. In this case, at the critical point, the CFT on both sides is the standard $c=1/2$ Ising CFT with the two sides communicating across the interface. The possible boundary conditions for Ising interfaces as well as their physics have been worked out  in \cite{Quella:2006de,Bachas:2013ora,Brehm:2015lja}. 

While in principle just a special case of a BCFT, the additional structure of an ICFT allows one to calculate several new observables. One famous example of a new dynamical observable that is only defined in ICFTs is the transmission coefficient \cite{Bachas:2001vj,Quella:2006de}: while in a BCFT any wave impinging on the boundary has to be reflected with 100\% probability to preserve unitarity, in an ICFT we can have a non-trivial reflection coefficient $R$ and a transmission coefficient $T$ where unitarity only demands $R+T=1$. This new observable gives rise to a very rich structure. An interface with $T=1$ is called topological as fusing multiple interfaces with $T=1$ always yields a new interface with $T=1$ irrespective of the distance \cite{Petkova:2000ip}. Non-topological interfaces can still be fused, but this requires some non-trivial renormalization group flow \cite{Bachas:2007td}. The other special case with $R=1$ corresponds to the case of two disconnected BCFTs that do not talk at all across the interface. 

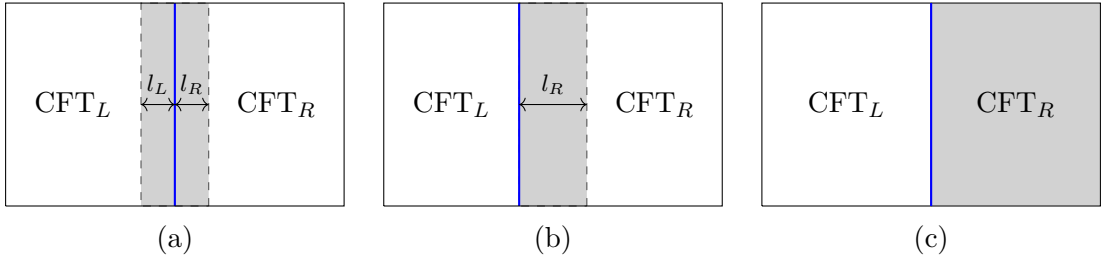
\begin{figure}[htb]
    \centering
    \begin{tikzpicture}[scale=0.9]
    \filldraw[dashed,fill=lightgray,opacity=0.7] (3,0)--(3,3)--(2,3)--(2,0)--(3,0);
    \draw[thick,blue] (2.5,0)--(2.5,3);
    \draw (0,0)--(5,0)--(5,3)--(0,3)--(0,0);
    \node at (1,1.5) {CFT$_L$};
    \node at (4,1.5) {CFT$_R$};
    \draw[<->] (2.5,1.5) -- (3,1.5);
    \node at (2.75,1.8) {\footnotesize $l_R$};
    \draw[<->] (2.5,1.5) -- (2,1.5);
    \node at (2.25,1.8) {\footnotesize $l_L$};
    \node at (2.5,-0.5) {(a)};
    \end{tikzpicture}
    \quad
    \begin{tikzpicture}[scale=0.9]
    \filldraw[dashed,fill=lightgray,opacity=0.7] (3,0)--(3,3)--(2,3)--(2,0)--(3,0);
    \draw[thick,blue] (2,0)--(2,3);
    \draw (0,0)--(5,0)--(5,3)--(0,3)--(0,0);
    \node at (1,1.5) {CFT$_L$};
    \node at (4,1.5) {CFT$_R$};
    \draw[<->] (2,1.5) -- (3,1.5);
    \node at (2.5,1.8) {\footnotesize $l_R$};
    \node at (2.5,-0.5) {(b)};
    \end{tikzpicture}
     \quad
    \begin{tikzpicture}[scale=0.9]
    \draw (0,0)--(2.5,0)--(2.5,3)--(0,3)--(0,0);
    \filldraw[draw=white,fill=lightgray,opacity=0.7] (2.5,0)--(5,0)--(5,3)--(2.5,3);
    \draw (2.5,0)--(5,0)--(5,3)--(2.5,3);
    \draw[thick,blue] (2.5,0)--(2.5,3);
    \node at (1.25,1.5) {CFT$_L$};
    \node at (3.75,1.5) {CFT$_R$};
    \node at (2.5,-0.5) {(c)};
    \end{tikzpicture}
    \caption{For all three figures, the gray areas bounded by the dashed lines are subsystems A's in the entanglement calculations. The blue lines are the interfaces. (a) A generic situation for case 1. The interface lives inside the spatial interval such that both $l_L$ and $l_R$ are nonzero. In this scenario, the interface is invisible to entanglement entropy at leading order in large $l$. The left and right boundaries of the interval both contribute $c/6$ to the prefactor of the leading term in entanglement entropy. (b) A special situation in case 1, where the interface coincides within left boundary of the interval such that $l_R=l$ and $l_L=0$. The right boundary of the interval still contributes $c/6$ to the prefactor of the leading term in entanglement entropy, while the contribution from left boundary gains an additional factor that depends on the transmission coefficient of the interface, which holographically corresponds to $e^{A_*}$. (c) Case 2. When the subsystem A extends to infinity on the right, the only contribution to the leading entanglement entropy comes from the left boundary of subsystem A.}
    \label{fig:setup}
\end{figure}

In contrast to this new dynamical probe specific to ICFTs, in this work we are interested in the entanglement structure in the ground states of an ICFT as encoded in the entanglement entropy (EE). In an interface theory, we have various options to calculate EEs that capture properties of the interface as laid out in detail in the review article \cite{Calabrese:2009qy}. There are two basic classes:
\begin{enumerate}
\item Trace out the degrees of freedom outside an interval of total length $l$ containing the defect. This is basically a standard entanglement entropy, albeit with an extra twist: the EE depends on the location of the defect within the interval. This gives rise to a two-parameter family of EEs characterized by $l_L$ and $l_R$, the lengths of the interval on the two sides of the defect. Clearly
\beq l = l_L + l_R . \eeq
Without loss of generality we take
\beq l_L \leq l_R .\eeq
Two special cases are
\begin{enumerate}
\item $l_L=l_R=l/2$, the symmetric interval 
\item $l_R=l$, $l_L=0$, the one-sided interval.
\end{enumerate}
Two examples of this case are shown in Figure \ref{fig:setup}(a), \ref{fig:setup}(b).
\item Trace out the degrees of freedom on one side of the defect, basically determining an inter-CFT entanglement, see Figure \ref{fig:setup}(c). The entanglement entropy is UV and IR divergent and so depends both on a UV regulator $\epsilon$ and an IR regulator $L$.
\end{enumerate}

There are a few properties of these EEs that have been uncovered in a ICFT where both sides of the interface are given by a CFT with central charge $c$. This is the case we will focus on for most of the paper. In section \ref{unequal} we will show that very similar results hold in ICFTs with unequal central charges $c_L$ and $c_R$ on the two sides of the interface. But for now let us summarize what is known about the case with equal central charge $c$, as for example reviewed in \cite{Calabrese:2009qy}:
\begin{itemize}
\item In case 1a) we have $S= \frac{2c}{6} \log l/\epsilon + \log g$, which is the standard BCFT result. For the symmetric interval $l_L=l_R$ case we can use the folding trick to reduce the system to the EE calculation in a BCFT with central charge $2c$ and an interval of size $l/2$. $g$ is a characteristic  constant of the defect  \cite{Cardy:1986gw,Affleck:1991tk}.
\item For the generic interval of case 1, we can not use folding to map to an EE calculation in a BCFT: After folding, the non-symmetric interval would correspond to a scenario where we are crucially making use of the product structure of the resulting BCFT. We are tracing out one set of degrees of freedom outside an interval of length $l_L$ around the boundary and the other set of degrees of freedom in an interval of size $l_R$. Not much seems to have been known about this case prior to this work.
\item For a completely one-sided interval, case 1b), it has been found \cite{Peschel_2005} that $S=\sigma_1 \log(l) + g$.  That is, the interface does not just give a constant contribution in $S$ but actually modifies the coefficient of the $log$ term. Clearly $\sigma_1=\frac{c}{3}$ in the case of a completely transparent interface (by which we mean no interface at all) where we reduce to the standard CFT result. Also clearly $\sigma_1=\frac{c}{6}$ in the case of two disconnected systems, as now we simply study an interval in a single BCFT.
\item In case 2) one also has $S= \frac{1}{2} \sigma_2 \log(L) + \log g $. This was first found in \cite{Sakai:2008tt} for an interface between two free compact bosons with a non-trivial jump in radius across the interface. It has since then been studied in a variety of CFTs, for example in \cite{Brehm:2015lja,Brehm:2015plf,Gutperle:2015hcv}. Overall, case 2 has been analyzed much more comprehensively since it is very amendable to path integral studies using the replica trick. 
There is an ambiguity in whether to include  a prefactor of 1/2 seemingly depending on whether one adds up contributions from both sides or not. In this work we follow the conventions of \cite{Gutperle:2015hcv} which is appropriate when calculating a genuine inter-CFT entanglement entropy \cite{Gutperle:2015kmw} and are to be contrasted with the original calculation in \cite{Sakai:2008tt} which included a contribution from both sides. While one finds $\frac{\sigma_2}{2} = \frac{c}{6}$, in the case of a completely transparent interface, this time one finds and $\sigma_2=0$ in the case of two disconnected systems. Clearly $\sigma_1 \neq \sigma_2$. 
\end{itemize}

In this work we are going to consider all these case, and in particular cover the generic case 1) with arbitary $l_L$ and $l_R$. All these EEs can in principle be calculated in holographic toy models. Holography postulates that some field theories have an equivalent description in terms of a higher-dimensional theory of gravity. A field theory is said to be holographic if its dual description is useful in that it can be solved using classical equations. Such field theories are rare, as they require a large central charge as well as a gap in the operator spectrum \cite{Heemskerk:2009pn,Hartman:2014oaa}. One well-studied example of a holographic ICFT is the so-called Janus solution, which was first worked out in the case of a 3+1 CFT in \cite{Bak:2003jk} and has been generalized to 1+1 dimensions in \cite{Bak:2007jm}. The 1+1 dimensional Janus CFT describes an interface between two CFTs that essentially describes $N$ compact bosons, where the radius of the compactification jumps across the interface. What makes this theory non-trivial is that this otherwise free theory is orbifolded by the symmetric group $S_N$ that permutes the $N$ bosons. As a result, the CFT really is a $\sigma$-model whose target space is $T^N/S_N$, where $T^N$ is an $N$-dimensional torus. This CFT has several marginal operators which correspond to the blowing-up of the orbifold singularity. While the CFT is tractable at the orbifold fixed point by standard CFT techniques, the holographic CFT corresponds to the limit of large blow-up parameters which is a regime of strong coupling. So, as it behoves for a duality, the regime in which gravity calculations are feasible is exactly the regime in which field theory calculations become challenging and vice versa. In this case of the 1+1 dimensional Janus ICFT, the EE has been explored in some special cases starting with the work of \cite{Azeyanagi:2007qj}. The holographic description of the special case 1a), the symmetric interval, was worked out and the equations for the general case 1 were laid out. The holographic description of case 2) was first presented in \cite{Gutperle:2015hcv}, both for the original Janus CFT as well as some of its supersymmetric generalizations.

In this work we will show that for holographic ICFTs, one can derive several very general features of the EE:
\begin{itemize}
\item For case 1), $S = \frac{c}{3} \log l + \log g_{\text{\text{eff}}}$ for {\it any} non-zero $l_L$ and $l_R$. The ratio of $l_L$ and $l_R$ only enters into some subleading term $\log g_{\text{\text{eff}}}(l_L/l_R)$, which is independent of $l$ and can be viewed as an effective interface entropy. This function seems to not be universal and depend on dynamical details of a given CFT.
\item The only non-trivial $\sigma$'s arise in the extreme limit of case 1b), that is $l_L=0$ and in case 2). These two seemingly different functions $\sigma_1$ and $\sigma_2$ are not independent but determined by a {\it single} geometric quantity, which we call $e^{A_*}$, the minimum warpfactor of the dual three-dimensional gravity. This quantity had already been found to dominate the case 2 in \cite{Gutperle:2015hcv}. Here we find that it also governs 1b), even though the two are not equivalent.
\item In terms of $e^{A_*}$ we have 
\beq \sigma_1 = \frac{c}{6} \left( 1+ e^{A_*} \right), \quad \frac{\sigma_2}{2} = \frac{c}{6} e^{A_*}. \label{upshot} \eeq
These expressions are consistent with the quoted results for the case of a transparent interface ($e^{A_*}=1$) and a completely disconnected interface ($e^{A_*}=0$). Together they imply a universal relation between $\sigma_1$ and $\sigma_2$:
\beq \sigma_1 = \frac{\sigma_2}{2} + \frac{c}{6} . \label{universal} \eeq
\end{itemize}

In this work we show that these results do not just hold in particular holographic models, but in fact hold in any holographic theory with a single holographic direction. The geometry of such models is characterized by a function $A(r)$ where $r$ is the coordinate along the holographic direction. $A(r)$ is known as the warpfactor. Given this generality, we suspect that these results do, in fact, hold in any CFT. If so, it would be reassuring to re-derive them using standard field theory techniques. Interestingly, while finishing this work, Ref. \cite{Kruthoff:2021vgv} appeared, where non-trivial $l_L$ and $l_R$ are considered for $2$d free fermions. So at least in the case of fermions, the results hold both at infinite as well as zero coupling, giving some credence to the belief that they may be, in fact, generic. Another reason that makes us suspect these results are universal is that they are dominated by UV physics. Furthermore, below we will provide an intuitive explanation for the universal relation \eqref{universal} which is independent of the existence of a holographic dual. 
While these arguments do not guarantee universality it at least makes it plausible. In contrast, should it turn out that these results are peculiar to holographic CFTs they could be used as a simple tool to determine whether a given ICFT could potentially have a holographic dual or not.

Our results can be summarized by a very simple set of rules for calculating the $l$-dependent leading term of the EE, which is explained in Figure \ref{fig:setup}. Since the entanglement is strongest at the endpoints of the interval, we associate all $\log l$-dependent terms to the boundaries of the interval,  such that each end of an interval somewhere away from the interface contributes $\frac{c}{6} \log l$. This gives the standard $\frac{c}{3} \log l$ for an interval (with two ends) in a CFT, and also explains why we get the same leading contribution for any interval including the interface with non-zero $l_L$ and $l_R$. A similar intuitive picture was used to understand the mutual information and multipartite information in field theory computations (see for example \cite{de2015entanglement}) and is also supported by the field-theoretic entanglement calculations when the subsystem A contains multiple disjoint intervals that are far apart \cite{Calabrese:2004eu}, where the leading term in entanglement entropy is a summation of leading terms for each interval. This has also been confirmed from a holographic \cite{Ryu:2006bv,Ryu:2006ef} perspective. Furthermore, we postulate that an interval boundary residing on the interface contributes $\frac{\sigma_2}{2} \log l$, accounting for the limited information transfer through the interval. A completely transparent interval has $\frac{\sigma_2}{2} = \frac{c}{6}$ and so is indistinguishable from a ``normal" interface boundary, whereas a true space boundary, or an interface with zero transmission, has $\sigma_2=0$. This recovers the familiar $\frac{c}{6} \log l$ for an interval in a BCFT as well as our universal relation \eqref{universal}.
That is, the difference between case 1b) and case 2) is that the former picks up an extra $c/6 \log l$ from the second interval boundary. 

Besides these general results on the leading contribution of EE, we also work out the non-universal aspects of the entanglement structure of ICFTs, i.e., the $g_{\text{eff}}$ function, for specific examples. The non-trivial dependence of $g_{\text{eff}}$ on the ratio $l_L/l_R$ encodes dynamical details of the CFT that is sensitive to the full functional form of the warpfactor $A(r)$. We do this in the Janus CFT as well as another popular ICFT, the subcritical RS brane \cite{Karch:2000gx,Karch:2000ct}. Unlike Janus, this is a ``bottom-up" model, meaning the dual CFT is not even known in principle. The subcritical RS brane simply is a toy-model for a putative CFT, presumably capturing essential aspects of generic holographic ICFTs. In this case the geometry is piecewise AdS$_3$, which makes the calculations very tractable.

The manuscript is organized as follows. In the next section we lay out the general holographic description of holographic CFTs and derive our main universal results. In Section 3 we study $g_{\text{eff}}$ in two specific models, the Randall-Sundrum (RS) braneworld and Janus cases. In Section 4 we discuss the case of unequal central charges on the two sides of the interface and, followed by the presentation of an alternate approach to the RS braneworld to verify our results in the appendix.

\section{Holographic ICFTs}
\label{sec:general}
\subsection{The Setup}

\begin{figure}[h]
\begin{centering}
\includegraphics[scale=0.79]{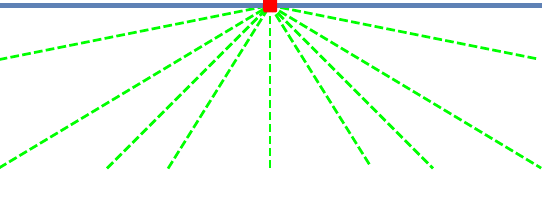}
\caption{Surfaces of constant $r$ in AdS$_3$, denoted by green dashed lines. The horizontal line on top is the asymptotic boundary of AdS$_3$, and the red dot is the interface. Lines of constant Poincar\'e Patch radial coordinate would correspond to horizontal lines. .\label{warpedsetup}}
\end{centering}
\end{figure}

Holographic ICFTs have  dual descriptions in terms of three dimensional gravity. To be UV completed within string theory, these gravitational duals typically involve 10d spacetimes with 7 compact dimensions. There are some examples in which the internal space is non-trivially fibered over the 3d spacetime of interest, as is for example the case in the gravitational duals for 3+1 dimensional supersymmetric Janus solutions of the type first developed in \cite{DHoker:2007zhm}. Here we limit ourselves to holographic ICFTs where the internal space is a genuine product factor in the metric  such that the spacetime has a good 2+1 dimensional description as is the case of the 1+1d Janus of \cite{Bak:2007jm}. The most general conformal defect spacetime in 2+1 dimensions takes the form
\beq ds^2 = e^{2A(r)} \frac{dx^2 - dt^2}{x^2} + dr^2. \label{metric} \eeq
This metric describes 2d slices whose overall size depends on the third dimension via the warpfactor $A(r)$. The metric on each slice is AdS$_2$, which is required in order to reproduce the symmetries of an ICFT.
The special case of AdS$_3$ itself is the dual of a completely transparent defect, that is a $1$+$1$d CFT with no defect at all, where we simply declare a random smooth point to be ``the defect". It corresponds to
\beq \label{emptywarp} e^A = \cosh(r)  .\eeq
In these coordinates 1+1 dimensional Minkowski space that is the boundary of AdS$_3$ is formally split into two halves, one living at $r \rightarrow + \infty$ and one at $r=-\infty$. They are nevertheless connected, with the missing interface that connects them corresponding to the asymptotic boundary of the AdS$_2$ space on each slice. Figure \ref{warpedsetup} depicts AdS$_3$ in these coordinates.
Technically speaking, in the coordinates of \eqref{metric} the field theory is living on two copies of AdS$_2$ meeting at their common boundary, which is conformally equivalent to $2$d Minkowski space. To interpret the results, it will often be helpful to go to the standard Poincar\'e coordinates on AdS$_3$, in which the boundary, located at $z=0$, is manifestly $2$d Minkowski space
\beq ds^2 = \frac{1}{z^2} (-dt^2 + dy^2 + dz^2 ). \label{poincare} \eeq
The two coordinate systems are related by\footnote{A third set of coordinates
we will sometimes employ uses a conformal coordinate for the warp product, $\cosh r = 1/\sin \mu$ (and hence $\cos \mu = \tanh r$) in terms of which the metric of AdS$_3$ reads
\beq ds^2 
= \frac{1}{\sin^2 \mu} \left( \frac{-dt^2 + dx^2}{x^2}  + d\mu^2 \right) .
\eeq
Using $\mu$ instead of $r$, the change of coordinates \eqref{coc} is just the standard change from Cartesian coordinates
$y$ and $z$  to spherical coordinates $x$ and $\mu$ on the plane, except for the fact that we chose the negative $y$-axis to be at $\mu=0$.}
\beq
\label{coc}
 z = \frac{x}{\cosh r}, \quad y = - x \tanh r.
\eeq

A general defect spacetime corresponds to a warpfactor with the following basic properties: (1) $r$ runs from $-\infty$ to $+ \infty$; (2) $e^A \sim \cosh(r - \delta r_{\pm})$ as $r \rightarrow\pm \infty$ for some constants $\delta r_{\pm}$. That is, far away from the defect we recover AdS$_3$; and (3) the minimal value of the warpfactor is $e^{A_*} \geq 0$. Unless stated otherwise, we parameterize the radial direction so that this minimal value is achieved at $r=0$. 

In Janus interfaces, $e^{A_*} <1$. As we shall see, $e^{A_*}$ corresponds to the quantity $\sigma_1$.  In empty AdS$_3$ we have $e^{A_*}=1$, this value corresponds to a transparent defect. In contrast, $e^{A_*}=0$ corresponds to a completely reflecting defect -- the spacetime literally falls apart into two disconnected halves. As we shall demonstrate in detail below, the EE's in case 1) for generic $l_L$ and $l_R$ are sensitive to the entire function $A(r)$, not just its minimal value $e^{A_*}$.

\subsection{Entanglement Entropies}

In order to calculate the entanglement entropy in this setup, we need to construct a minimal surface in the 3d spacetime, ending on the locations that separate the two entangling subsystems, which is the so-called Ryu-Takayanagi (RT) surface \cite{Ryu:2006bv}.

\subsubsection{Case 2}
Let us start with discussing the case 2) of the inter-CFT entanglement entropy. That is, we are looking for an RT surface which ends on the interface at $x=0$ without any other endpoint on the boundary. This is exactly the case studied in \cite{Gutperle:2015hcv} where it was found that the correct RT surface corresponds to setting $r=0$,  which is simply the slice located at the minimum warpfactor. In figure \ref{warpedsetup} it corresponds to the central slice, reaching vertically down from the interface. To see this is indeed a minimal surface, we parametrize the RT surfaces as $t=0$ and $r(x)$. In this case the area functional (which we will refer to as the Lagrangian for the RT surface) becomes
\beq {\cal L} = \sqrt{(r')^2 +  \frac{e^{2A}}{x^2} }. \eeq
The equation of motion is indeed satisfied with $r'=0$ as long as $A'=0$,  when the warpfactor is at its minimum. In this case we have
\beq S = \frac{1}{4G}  e^{A_{*}} \int_{\epsilon}^{L} d x =  \frac{e^{A_{*}}}{4G} \log L/\epsilon, \eeq
where $G$ is the three-dimensional Newton's constant. This indeed has the expected form with
\beq \frac{\sigma_2}{2}= \frac{e^{A_{*}}}{4G} = \frac{c}{6} e^{A_*} . \label{sigma2} \eeq
In the last step we have used the Brown-Henneaux relation \cite{Brown:1986nw} for the central charge, where the AdS$_3$ radius is chosen to be one:
\beq \label{brh} G = \frac{3}{2c}.\eeq

\subsubsection{Case 1}

For case 1), we are looking for a minimal area surface at $t=0$ that reaches the boundaries at $x=l_R$ when $r \rightarrow + \infty$ and at $x=l_L$ when $r \rightarrow - \infty$.
In this parameterization the Lagrangian 
becomes
\beq {\cal L } = \sqrt{1+ e^{2A} \frac{(x')^2}{x^2} }. \eeq
The scale isometry of AdS$_2$ on the slice maps to a symmetry $x \rightarrow \lambda x$ of the Lagrangian, the corresponding Noether charge tells us that
\beq \frac{e^{2A} x'}{\sqrt{x^2 + e^{2A} (x')^2}} = c_s, \eeq
where $c_s$ is an integration constant. Solving for $x'$ we find
\beq
\label{solution}
\frac{x'}{x} = \pm \frac{c_s e^{-A}}{\sqrt{e^{2A}-c_s^2}} . \eeq
From this solution we can immediately read off the range of $c_s$:
\beq 0 \leq c_s \leq e^{A_*} .\eeq
Since our solution has a $\pm$ ambiguity anyway, nothing is lost by restricting to positive $c_s$. Interface-crossing intervals need to run over the entire range of $-\infty < r < \infty$, so all possible values of $e^A$ are realized and for the square root to remain real, we need to restrict ourselves to  $c_s \leq e^{A_*}$.

$c_s=0$ corresponds to $x=l_R=const.$ This is the symmetric case 1a) with $l_L=l_R$. Since $1/\sqrt{1-c_s^2 e^{-2A}}$ is always positive, the solution with the + sign in \eqref{solution} corresponds to $l_R \geq l_L$ and, without loss of generality, we will limit ourselves to that case. The special limiting value of $c_s = e^{A_*}$  corresponds to the case 1b) with $l_L=0$. To see this, note that in this special case the solution is singular at $r=0$ unless $x=0$ at this point. That is what the $l_L=0$ solution should look like: we hit the defect at $x=0$ exactly at the central $r=0$ slice. Note that this case is really special in this respect: only for $c_s=e^{A_*}$  will $r$ not cover its entire range from $-\infty$ to $\infty$ but instead truncate at $r=0$. The limit of $l_L \rightarrow 0$ is not smooth: even for infinitesimally small $l_L$ will $r$ run over its entire range. The upshot of this discussion is that the parameter $c_s$ dials $l_L$ from $l_L=l_R$ at $c_s=0$ to $l_L=0$ when $c_s$ reaches its maximum value.

The on-shell Lagrangian evaluated on the solution is given by 
\beq
\label{onshelll}
{\cal L} = \frac{1}{\sqrt{1 - c_s^2 e^{-2A}}}.
\eeq
To evaluate the corresponding area, we need to discuss how the integral of ${\cal L}$ is regulated. To understand this, let us first study the case empty AdS$_3$ in this coordinate system, where the answer is known. Since the warpfactor is demanded to asymptotically become that of the AdS$_3$, the regulating procedure we use for general $e^A$ will be identical to the one in AdS$_3$. Plugging in $e^A = \cosh r$ into our solution \eqref{solution}, we find that
for $0 < c_s < e^{A_*}=1$ we can integrate
\beq \label{adssolution} \frac{x}{x_0} = e^{\tanh^{-1} \frac{c_s \sinh r}{\sqrt{\cosh^2 r - c_s^2}}} .\eeq

For the special case 1a) with $c_s=0$, we recover the constant $x=x_0=l_R$. For $0<c_s < 1$, the solution corresponds to\footnote{Note the following identity for $\tanh^{-1}$:
\beq \tanh^{-1} x = \log \frac{\sqrt{1+x}}{\sqrt{1-x}} . \label{footidentity} \eeq}
\beq 
l_R = x_0 e^{\tanh^{-1} c_s} = x_0 \frac{\sqrt{1+c_s}}{\sqrt{1-c_s}}, \quad l_L = x_0 e^{- \tanh^{-1} c_s}
= x_0 \frac{\sqrt{1-c_s}}{\sqrt{1+c_s}}. \label{alpharelation} 
\eeq
For the case 1b) with $c_s = e^{A_*}=1$, $x'/x$ integrates to
\beq x = x_0 \tanh r \label{limitsolution} \eeq
instead. Clearly $l_R = x_0$ and $l_L=0$ in this case.

The on-shell Lagrangian for this AdS$_3$ background simplifies to
\beq {\cal L} = \frac{\cosh r}{\sqrt{\cosh^2 r - c_s ^2}} . \label{onshellads} \eeq
Let us first look at the special case of $c_s=0$. In this case ${\cal L}=1$. If we regulate the area by truncating the integral both a large positive  $r$ by $r=r_c^+$ and at large negative $r$ by $r=-r_c^-$, the regulated area simply becomes
\beq {\cal A} = r_c^+ + r_c^- .\eeq
While this is a stunningly simple answer, at the face of it this appears to disagree with what we know the right answer should be, ${\cal A} = 2 \log(2 l_R/\epsilon)$, where $\epsilon$ is the UV cutoff and $l=2l_R$ the total length of the interval. To understand this discrepancy, note that in the metric we are working with, the 1+1 dimensional spacetime our field theory is living on is AdS$_2$. The important thing to note is that a position-independent cutoff in AdS$_2$ corresponds to a position-dependent cutoff if we conformally transform to 2d Minkowski space, and vice versa. This can easily be made quantitative by noting that in order to relate our metric to the standard Poincar\'e patch metric on AdS$_3$, \eqref{poincare}, the corresponding change of variables \eqref{coc}, near $r\rightarrow \infty$ reads
\beq \frac{e^r}{2} =\frac{x}{z}  \label{cutoff} . \eeq
Correspondingly, a standard position-independent UV cutoff with $z=\epsilon$ for 2d Minkowski space corresponds to a cutoff at large positive $r$ at
\beq r_c^+ = \log \frac{2 l_R}{\epsilon} \label{reg} .\eeq
At large negative $r$ we similarly get
\beq 
r_c^- = \log \frac{2 l_L}{\epsilon} \label{reg2}. 
\eeq
With this, our answer for the area in this $l_L = l_R$ case indeed becomes
\beq {\cal A} = r_c^+ + r_c^- = 2 \log (2 l_R/\epsilon ). \eeq
So in the $r$ coordinates the correct regularization procedure is given by \eqref{reg} for the cutoff at large positive $r$, and similar at large negative $r$ by \eqref{reg2}.

Next, let us look at the case of generic $0 < c_s < 1$ in the empty AdS$_3$ background. The on-shell action \eqref{onshellads} integrates to
\beq {\cal A} = \label{integratedaction} \left . \tanh^{-1} \frac{\sinh r}{\sqrt{\cosh^2 r - c_s^2}} \right |_{-r_c^-}^{r_c^+}  =
 \log (2 l_R/\epsilon) +  \log (2 l_L/\epsilon)   - \log (1-c_s^2),\eeq
where we used \eqref{footidentity} to expand $\tanh^{-1}$ at large $r_c$.
What we know this answer {\it should} be is, again, ${\cal A}=2 \log l/\epsilon$ where $l=l_R + l_L$. To confirm this is the case let us make the following substitution:
\beq \label{definealpha} l_L = \alpha l, \quad l_R = (1- \alpha) l, \eeq
where $0 \leq \alpha < 1/2$. This allows us to identify
\beq   {\cal A}  = 2 \log l/\epsilon   + \log \frac{4 \alpha (1- \alpha) }{1-c_s^2} . \eeq
Last but not least, we need the relation between $\alpha$ and $c_s$. From \eqref{alpharelation} we find
\beq 
\frac{l_L}{l_R} =  \frac{\alpha}{1 - \alpha} = \frac{1-c_s}{1+c_s} \quad \Rightarrow \quad
\label{alpharelation'} \alpha =  \frac{1-c_s}{2}. 
\eeq
With this 
\beq 4 \alpha (1- \alpha) = 1 - c_s^2 \quad 
\Leftrightarrow \quad c_s = 1- 2 \alpha,
\label{csalpharelation}
\eeq
and so indeed
\beq {\cal A} = 2 \log l/\epsilon . \label{eq:leading}\eeq
Our cutoff procedure is consistent and reproduces known results!

\subsubsection{Universality of the \texorpdfstring{$\log l$}{} term}

Last but not least, let us derive the universal results for the coefficient of the $\log l$ term for general warpfactor advertised in the introduction. This requires considering separately the $l_L >0$ case and the $l_L=0$ case, corresponding to $c_s < e^{A_*}$ and $c_s=e^{A_*}$ respectively. Let us start with the former. In this case, we aim to prove that the coefficient of the $\log l$ term in $S_{EE}$ is universally $c/3$, independent of $l_L/l_R$. Since $c_s < e^{A_*}$, $r$ runs over the entire range from $-\infty$ to $+ \infty$. While one needs to commit to a particular form of $A(r)$ in order to solve the full entanglement entropy,  its $l$-dependence can be extracted quite generally.

To do so, it is easiest to not calculate the area ${\cal A}$ itself but rather its variation as we change $l_L$ and $l_R$. For this, one should first of all note that the on-shell Lagrangian \eqref{onshelll} is determined in terms of the warpfactor itself and is completely independent of $l_L$ and $l_R$. So the entire dependence on $l_L$ and $l_R$ comes from the cut-off. The cutoffs depend on $l_{L/R}$ via \eqref{reg} and \eqref{reg2}. This yields
\beq \label{deltaa}
\delta {\cal A}  = \left . {\cal L} \right |_{r=r_c^+} \frac{\delta r_c^+}{\delta l_{R} } \delta l_R  - \left . {\cal L} \right |_{r=-r_c^-} \frac{\delta (-r_c^-)}{\delta l_{L} } \delta l_L . \eeq
In the large $|r_c|$ limit, the on-shell Lagrangian \eqref{onshelll} just goes to 1. Furthermore from \eqref{reg} and \eqref{reg2} we have
\beq
\frac{\delta r_c}{\delta l_{L/R}} = \frac{1}{l_{L/R}}.
\eeq
So lo and behold we arrive at
\beq
\frac{\delta {\cal A}}{\delta_{l_{R/L}}}  = \frac{1  }{l_{R/L}},  \eeq
and hence
\beq {\cal A} =  \log l_L/\epsilon +  \log l_R/\epsilon  + const.  = 2 \log l/\epsilon + const., \label{eq:universal_log}\eeq
corresponding to a universal $c/3 \log l$ term in $S_{EE}$ as advertised. (Note that the two constants in the equation above are different.)

Next let us turn to the special case of 1b), with $c_s=e^{A_*}$. In this case $l_L=0$ and $l=l_R$. This is the only place where we can potentially find a non-universal coefficient of the $\log l$ term. Let us re-do the analysis of the $l_{L/R}$-dependence of the entanglement entropy in this case. The on-shell Lagrangian is the same \eqref{onshelll} as before, but this time the endpoints of the $r$-integration are naively 0 and $r_c^+$. Recall that we chose the origin of $r$ so that $r=0$ is where the warpfactor reaches its minimal value $e^{A_*}$. 

Clearly $\log l$ dependence will arise near $r_c^+$ as before. What is a little less obvious is that the integral in fact also diverges at $r=0$, so we need to regulate this contribution as well, leading to further $\log$ $l$-dependence. Despite $r$ remaining finite, the integral diverges as we are reaching $x=0$ as $r \rightarrow 0$ for this special solution. This still means the RT surface hits the boundary (right on the interface) and so this is a standard UV divergence. On a constant $r$ slice, the spatial metric reads $ds^2 = e^{2A} dx^2/x^2$, so a cutoff at a constant Poincar\'e coordinate $z=\epsilon$ amounts to
\beq e^{A_R} \epsilon = x, \label{trivialcut} \eeq
where $e^{A_R}=e^{A(r_c^0)}$ is the warpfactor at the cutoff slice. Since the warpfactor is minimal at $r=0$, we have $A'(0)=0$ and so
\beq e^{A_R} = e^{A_*} + {\cal O} ((r_c^0)^2), \label{eq:AR}\eeq
as the linear term in the Taylor expansion vanishes. Revisiting the variation in area from \eqref{deltaa}, we this time get
\beq \label{deltaa2}
\frac{\delta {\cal A}}{\delta l}  = \left . {\cal L} \right |_{r=r_c^+} \frac{\delta r_c^+}{\delta l }   - \left . {\cal L} \right |_{r=r_c^0} \frac{\delta r_c^0}{\delta l}  . \eeq
The term at $r_c^+$ gives us $1/l$ just as before. To evaluate the second term, first note that from \eqref{onshelll} with $c_s=e^{A_*}$ we have
\beq \left . {\cal L} \right |_{r=r_c^0} = \frac{e^{A_R}}{\sqrt{e^{2 A_R} - e^{2 A_*}}}  =
\frac{e^{A_*}}{\sqrt{e^{2 A_R} - e^{2 A_*}}} + {\cal O}(r_c^0) .\eeq
 Obtaining $\delta r_c^0/\delta l$ requires a little more work. While the on-shell action turned out to be independent of $l_{L/R}$, the details of the profile $x(r)$ do depend on $l$. As we will see in detail in our examples, different boundary conditions at the $r \rightarrow \infty$ end correspond to different RT surfaces. But this means that when imposing \eqref{trivialcut} we need access to the full solution $x(r)$ in order to understand the full $r_c^0$ as a function of $l$. This seems to require us doing the integral in \eqref{solution}, which can not be done in closed form for general warpfactor. Fortunately we do not need $r_c^0$ itself, but only its variation. This we can get a handle on. Reading \eqref{solution} as an equation for $(\log x)'$ and integrating from $r_c^0$, where $x= e^{A_R} \epsilon$, to $r=r_c^+$, where $x=l$, we can write
\beq \log \left ( \frac{l}{e^{A_R} \epsilon} \right ) = \int_{r_c^0}^{r_c^+} \frac{e^{A_*-A}}{\sqrt{e^{2A} - e^{2A_{*}}}}. \eeq
Taking an $l$ derivative on both sides this implies
\beq \frac{1}{l} = \left . \frac{e^{A_*-A}}{\sqrt{e^{2A} - e^{2A_{*}}}} \right |_{r_c^+} \frac{\delta r_c^+}{\delta l}
- \left . \frac{e^{A_*-A}}{\sqrt{e^{2A} - e^{2A_{*}}}} \right |_{r_c^0}  \frac{\delta r_c^0} {\delta l}. \eeq
The first term on the right hand side vanishes as the warpfactor goes to zero at large $r$. So we can solve for $\delta r_c^0/\delta l$ as
\beq \frac{\delta r_c^0}{\delta l} = - \frac{1}{l} \frac{\sqrt{e^{2A_R} - e^{2 A_*}} }{e^{A_* - A_R}} =
- \frac{1}{l} \sqrt{e^{2A_R} - e^{2 A_*}} + {\cal O}(r_c^0) \eeq
Putting these results back into the variation of the area \eqref{deltaa2} and dropping terms that vanish in the $r_c^0 \rightarrow 0$ limit we find
\beq \frac{\delta {\cal A}}{\delta l} = \frac{1}{l} + \frac{e^{A_R}}{\sqrt{e^{2 A_R} - e^{2 A_*}}}\sqrt{e^{2A_R} - e^{2 A_*}}  \frac{1}{l} = \frac{1 + e^{A_*}}{l}.\eeq
This corresponds to a $\sigma \log{l}$ term in the entanglement entropy with
\beq
\sigma_1 = \frac{c}{6} (1 + e^{A_*}). \label{sigma1}
\eeq
Together with our previous result \eqref{sigma2}, this indeed implies the universal relation \eqref{universal} advertised in the introduction. The way this result was derived also strongly supports the model we outlined in the introduction where we associated $\log l$ terms with the ends of the interval and assemble the general case from contributions from the two ends of the interval. This is exactly what happens in the gravitational solutions. $\log l$-dependence only arises from the cutoffs at either end of the RT surface. A UV divergence at large $|r|$ but finite $x$ gives a $\frac{c}{6} \log l$, whereas a UV divergence at $x=0$ gives $\frac{\sigma_2}{2} \log l$.

\section{Examples}
\subsection{The RS braneworld}

One special example of a known warpfactor $A(r)$ is the subcritical Randall-Sundrum (RS) braneworld \cite{Karch:2000gx,Karch:2000ct}. An RS braneworld is a very simple model in general relativity -- the action describing the bulk theory is that of Einstein gravity with a negative cosmological constant coupled to matter.  The matter is a brane -- a thin relativistic sheet with constant energy density per unit volume, characterized by a single dimensionful quantity, its tension. As long as the tension remains below a certain critical value, the dynamics of the brane demands that it intersects the boundary along a time-like defect, with the spacetimes on both sides of the defect being empty AdS. This way, the spacetime with the brane can naturally be interpreted \cite{Karch:2000gx,Karch:2000ct,Takayanagi:2011zk} as a holographic dual of an ICFT if we take the spacetime as it is, or a BCFT if we impose an additional orbifold condition which identifies the two halves of the spacetime.\footnote{As recently pointed out in \cite{Bachas:2021fqo}, a different route to reach a BCFT from the RS braneworld is to let the cosmological constant jump across the brane together with its geometric backreaction which leads to a jump in extrinsic curvature. This basically means the brane is charged under a  3-form flux in the 3d spacetime. A BCFT corresponds to ``no spacetime", that is zero curvature radius, on one side of the brane. This requires an infinite tension. What appears as the tension in \cite{Takayanagi:2011zk} is a subleading term in this interpretation.} While being a simple solution of general relativity, these braneworlds, as they stand, do not arise as a low energy limit of string theory and so, while one can interpret their physics in terms of a dual CFT, the dual CFT can not be explicitly constructed.\footnote{One notable exception is the case of the $c=1/2$ Ising model, where we argued in our earlier work \cite{Karch:2020flx} that the full Ising model is dual to pure Einstein gravity plus brane, albeit in a highly quantum regime.} Such low energy toy models are known as bottom-up constructions.

For our purposes, we can simply think of the RS braneworld as a special case of a warpfactor 
\beq e^A = \cosh(|r|-r_*) \label{rswarp} .\eeq
That is, the solution is piecewise defined for positive and negative $r$, where $r=0$ is the location of the brane and its distance to the turnaround at $r=r_*$ is fixed uniquely by the tension $T$.
The warpfactor is continuous, but its first derivative has a discontinuity -- there is a non-trivial jump in the extrinsic curvature due to the stress energy of the brane. From the study of the symmetric interval $l_L=l_R$ case, case 1a), it is known that the tension, and hence $r_*$, encodes the interface entropy  $\log g_{\text{eff}}$. The three are related as \cite{Takayanagi:2011zk}
\beq
\label{gfactor}
\log g = \frac{2 r_*}{4 G} = \frac{c r_*}{3} = \frac{c}{3} \tanh^{-1} \frac{T}{2}.
\eeq
We will have to recover this result when thinking of RS as a special case of our general setup.

Note that this case is slightly different from our previous discussion in that the minimum of the warpfactor is now reached at $r=\pm r_*$ where $e^{A_*}=1$. If we wanted to insist that the minimum warpfactor is obtained at $r=0$, we could always shift the radial coordinate so that one of the two degenerate minima is sitting at $r=0$. But in this special case ,it is more convenient to have the brane at $r=0$ with the two minima symmetric around it. It is still true that the parameter $c_s$ is bounded above by 1, the minimal value of $e^A$, to make sure the solution \eqref{solution} for $x'/x$ is real.

There have been quite a few recent studies of RT surfaces in RS spacetimes in the context of entanglement islands, for example \cite{Almheiri:2019psy,Chen:2020uac,Geng:2020fxl,Geng:2021wcq}. In this context, one usually first solves the RT surface in the un-deformed spacetime away from the brane and then imposes a non-trivial gluing condition for the RT surface across the brane. We will show in Appendix \ref{app} that the results we find by simply viewing the RS warpfactor as an example of our general formulae agree with the cut-and-paste method for the RS spacetime.

Plugging the specific warpfactor \eqref{rswarp} into
our general solution \eqref{solution}, we essentially obtain once again the empty AdS$_3$ RT surface \eqref{adssolution}, but with a shifted argument. For the derivative of the embedding function $x(r)$ we immediately get
\beq 
\label{rssolution} 
\frac{x'}{x} =  \frac{c_s \sinh (|r|-r_*) }{\sqrt{\cosh^2(|r|-r_*) - c_s^2}},
\eeq
where we chose the sign in \eqref{solution} so that $l_R \geq l_L$ as in \eqref{adssolution}.
To obtain $x(r)$ we integrate piecewise for $r>0$ and $r<0$. After freely choosing the integration constant at $r>0$ the integration constant at $r<0$ is fixed by requiring continuity of the function $x(r)$. We find 
\beq 
\label{rssolutionintegrated} 
x = 
\left \{ \begin{array}{lll}
x_0 \, e^{\tanh^{-1} \frac{c_s \sinh (r-r_*)}{\sqrt{\cosh^2 (r-r_*) - c_s^2}}}, & \mbox{ for } & r>0 \cr
\cr
\tilde{x}_0 \, e^{\tanh^{-1} \frac{c_s \sinh (r+r_*)}{\sqrt{\cosh^2 (r+r_*) - c_s^2}}}, & \mbox{ for } & r<0
\end{array} \right .
\eeq
with 
\beq
\frac{x_0}{\tilde{x}_0} = e^{2 \tanh^{-1} \frac{c_s \sinh (r_*)}{\sqrt{\cosh^2 (r_*) - c_s^2}}}.
\label{x0relation}
\eeq
The relation
\eqref{alpharelation} between the integration constants $x_0$, $c_s$ and $l_L$, $l_R$ correspondingly also has to be modified to
\beq 
l_R = x_0 e^{\tanh^{-1} c_s} = x_0 \frac{\sqrt{1+c_s}}{\sqrt{1-c_s}}, \quad l_L = \tilde{x}_0 e^{- \tanh^{-1} c_s}
= \tilde{x}_0 \frac{\sqrt{1-c_s}}{\sqrt{1+c_s}}, \label{newalpharelation} 
\eeq
that is for $l_L$, $\tilde{x}_0$ given by \eqref{x0relation} replaces $x_0$ from \eqref{alpharelation}.
With this the relation between $c_s$ and $\alpha$ originally in \eqref{csalpharelation} gets modified to
\beq
c_s = \frac{(\tilde{x}_0/x_0) (\alpha-1) + \alpha}{(\tilde{x}_0/x_0) (\alpha-1) - \alpha} \quad
\Leftrightarrow \quad 
c_s = \frac{(1-2 \alpha) \cosh r_* \left(\cosh r_* - 2 \sinh r_* \sqrt{(1-\alpha) \alpha}\right)}{\cosh^2 r_* - 4 \sinh^2 r_* \, \alpha (1-\alpha)} .
\label{newcsalpharelation}
\eeq

Following the steps that led to \eqref{integratedaction}, we find
\begin{eqnarray}
\nonumber
{\cal A} &=&  \left . \tanh^{-1} \frac{\sinh \tilde{r}}{\sqrt{\cosh^2 \tilde{r} - c_s^2}} \right |_{-r_c^-+r_*}^{r_*}
+  \left . \tanh^{-1} \frac{\sinh \tilde{r}}{\sqrt{\cosh^2 \tilde{r} - c_s^2}} \right |_{-r_*}^{r_c^+-r_*}
\label{eq:3_4}
\end{eqnarray}
Here $\tilde{r}=r\pm r_*$ for negative (positive) $r$.\footnote{Note the difference from the cutoff procedures in \eqref{reg2}.} Naively, one might think that the shift of the uppper/lower integration boundary gives an extra contribution, but this is just a part of the cutoff procedure. At large positive $\tilde{r}$, we have in analogy with \eqref{cutoff}, 
\beq \frac{e^{\tilde{r}}}{2} =\frac{x}{z}.  \eeq
So if we want the cutoff slice at $\tilde{r} = r_c -r_*$ to correspond to a cutoff at $z=\epsilon$, we need that
\beq
\tilde{r}_c^+ = r_c^+ - r_* = 2 \log(2 l_R/\epsilon).
\eeq
and the cutoff contributions are exactly the same as  in \eqref{eq:leading}. The only new contribution comes from the central part away from the IR truncations,
\beq {\cal A} = 2 \log \frac{l_L + l_R}{\epsilon} + \Delta {\cal A}, \eeq
where
\beq
\Delta {\cal A} = 
\left . \tanh^{-1} \frac{\sinh \tilde{r}}{\sqrt{\cosh^2 \tilde{r} - c_s^2}} \right |_{-r_*}^{r_*}
= 2 \tanh^{-1} \frac{\sinh r_*}{\sqrt{\cosh^2 r_* - c_s^2}}.
\eeq
Here $c_s$ should be thought of as being a function of $\alpha$ given by \eqref{newcsalpharelation}. 
Recall from \eqref{definealpha} that $\alpha$ varies from 0 to 1/2, where $\alpha=1/2$ corresponds to the case $l_L=l_R=l/2$ and $\alpha=0$ to $l_L=0$, $l_R=1$. Our result verifies that indeed for any non-vanishing $\alpha$, the overall dependence on the length of the interval is {\it always} ${\cal A} = 2 \log l/\epsilon$ and hence 
\beq
S = \frac{c}{3} \log l/\epsilon + \log g_{\text{eff}}(\alpha) .
\eeq
with
\beq \log g_{\text{eff}} = \frac{c}{6} \Delta {\cal A} .\eeq
Non-vanishing $\alpha$ only affects the length-independent ``interface entropy".
$g_{\text{eff}}$ is a highly non-trivial function of $\alpha$. Furthermore, it seems to be non-universal in that it is sensitive to the details of the full $e^A$ in \eqref{rswarp} and not just to its minimal value.
$\log g_{\text{eff}}$  diverges as $\alpha \rightarrow 0$, consistent with the observation that in this limit we will change the UV-divergent logarithmic term and not just the $l$-independent constant.  

This special case of $l_L = \alpha l = 0$ once again has to be studied separately. It corresponds to the choice $c_s=1$. Recall that in general the maximal value for $c_s$ is set by the minimum value of $e^A$. What is special in the RS solution is that this minimum value is not obtained at $r=0$, the location of the brane, but instead at $r= \pm r_s$ where $e^A=1$. For $c_s <1$, we have $x(r)>0$ for all $r$ and the RT surface goes from $r=-\infty$ all the way to $r=+\infty$. This is the same behavior we've seen for generic warpfactor. We've also seen that in the special case of $c_s=1$, the RT surface has to reach $x=0$ exactly at the location of the minimum warpfactor, $r=r_s$ in this case. So the effective range of $r$ this time is only $r_* \leq r < \infty$. As we did before for empty AdS, the general solution as written in \eqref{rssolution} becomes singular in this case and we should instead generalize \eqref{limitsolution} which now reads
\beq x = l_R \tanh(r-r_*). \label{trivialsolution} \eeq
The on-shell action \eqref{onshelll} simply becomes ${\cal L} = \coth(r-r_*)$, and so formally
\beq \label{naive} {\cal A}_{\text{naive}} = \log (\sinh \tilde{r}) \big{|}_{r_*}^{\log (2l_R)/\epsilon} . \eeq
As we discussed when looking at the general case, this naive answer diverges at $r=r_*$ as the integrand diverges. The regulator condition \eqref{trivialcut}, which was chosen to have a position-independent cut off in Minkowski space, instructs us to cut off the integral at $\tanh r_c^0 = \epsilon/l_R$. So we find the full answer is
\beq {\cal A} = 2 \log l_R/\epsilon \eeq
That is, we simply get the same answers as in empty space without any contribution from the defect. This feature is extremely peculiar to RS -- the metric is unchanged from empty AdS$_3$ for all $r>0$ away from the brane, so the $\alpha=0$ RT surface simply does not know it is there at all.

\subsection{Janus}

In this section, we study another example called the Janus solution, which was originally constructed by \cite{Bak:2003jk}.\footnote{Study of its pure 4d field-theoretic aspects includes \cite{Gaiotto:2008sd,Ganor:2014pha,Ganor:2019nnv}.} Later in \cite{Bak:2007jm}, the Janus dilatonic deformation in type IIB supergravity of the AdS$_3\times S^3\times T^4$  vacuum was studied. The 10-dimensional metric is
\beq
ds^2_{\text{IIB}}=e^{\phi/2}(ds^2_{(3)}+d\Omega_3^2)+e^{-\phi/2}ds_{T^4}^2,
\eeq
where the 3d part of the theory is that of the Einstein-Hilbert action plus a dilaton field $\phi$, with the metric given by \cite{Freedman:2003ax}:
\beq 
ds^2=dr^2+ \frac{1}{2} \left(1+\sqrt{1-2\gamma^2}\cosh 2r\right)~ds_{AdS_2}^2.
\eeq
Here $|\gamma|\leq 1/\sqrt{2}$ and is related to the asymptotic values of the dilaton $\phi_\pm(r\rightarrow\infty)$ as follows
\beq \phi_\pm(r\rightarrow\infty)=\phi_0\pm\frac{1}{2\sqrt{2}}\log \left(\frac{1+\sqrt{2}\gamma}{1-\sqrt{2}\gamma}\right).\eeq
When $\gamma=0$, the dilaton field is a constant $\phi_0$, and the metric simply reduces to that of the empty AdS$_3$ spacetime. When $\gamma=1/\sqrt{2}$, the spacetime is AdS$_2\times\mathbb{R}$. 
We define $\xi\equiv\sqrt{1-2\gamma^2}$ for convenience, such that the warp factor defined in \eqref{metric} is $e^{A(r)}=\sqrt{(1+\xi\cosh 2r)/2}$. The integration constant $c_s$ in \eqref{solution} is then bounded by $0\leq c_s\leq \sqrt{(1+\xi)/2}$. We will first study the generic $c_s$ satisfying $0< c_s< \sqrt{(1+\xi)/2}$ and discuss the special cases of $c_s=0$ and $c_s=\sqrt{(1+\xi)/2}$ later in this subsection. For a general $c_s$, equation \eqref{solution} with the Janus warpfactor then leads to the following solution of $x(r)$:
\beq
\frac{x}{x_0}=\exp\left[-\frac{r}{|r|} \frac{2 c_s }{(\xi-1)^{1/2}(\xi+1-2c_s^2)^{1/2}} F(c_s,\xi,r)\right],
\eeq
where
\beq
F(c_s,\xi,r)\equiv F\left(\arcsin \frac{(2c_s^2-\xi-1)^{1/2}(\xi+\xi\coth^2r+\text{csch}^2r)^{1/2}}{2\xi^{1/2} c_s}\Bigg{|} \frac{4\xi c_s^2}{(\xi-1)(-\xi-1+2c_s^2)}\right),
\eeq
and $F(\phi|m)$ is the incomplete elliptic integral of the first kind, defined as $F(\phi|m)=\int_0^\phi d\theta (1-m\sin^2\theta)^{-1/2}$ for $m\in\mathbb{R}$ and $\phi\in(-\pi/2,\pi/2).$ Taking the $r\rightarrow \pm \infty$ limits of the equation above, one can then find 
\beq
l_{R/L}= x_0 \exp \left\{\pm\left[\frac{2 c_s }{(\xi-1)^{1/2}(\xi+1-2c_s^2)^{1/2}} F(c_s,\xi, \infty)\right]\right\},
\eeq
where
\beq
F(c_s,\xi,\infty) \equiv F\left(\arcsin \left(1-\frac{1+\xi}{2c_s^2}\right)^{1/2}\Bigg{|} \frac{4\xi c_s^2}{(\xi-1)(-\xi-1+2c_s^2)}\right).
\eeq

To find the regulated area ${\cal A}$, we need to integrate over the on-shell Lagrangian, which is again \eqref{onshelll} with the Janus warpfactor. Now we describe the appropriate cutoff procedure in analogy with \eqref{cutoff}. Recall that asymptotically, the AdS$_3$ warpfactor approaches $e^A \sim e^r/2$, giving rise to \eqref{cutoff} when changing coordinates back to Poincar\'e $z=\epsilon$. The asymptotic form of the Janus warpfactor, in comparison, approaches $e^A \sim e^r\xi^{1/2}/2$.  
Correspondingly, \eqref{cutoff} needs to be replaced by
\beq
\frac{e^r \xi^{1/2}}{2}= \frac{x}{z}.
\eeq
Furthermore, instead of \eqref{reg} and \eqref{reg2}, we now have 
\beq r_c^{\pm} = \log \frac{2 l_{R/L}}{\epsilon}-\log \xi^{1/2} . 
\label{eq:janus_cutoff}
\eeq

\begin{figure}[h]
\centering
\includegraphics[scale=0.6]{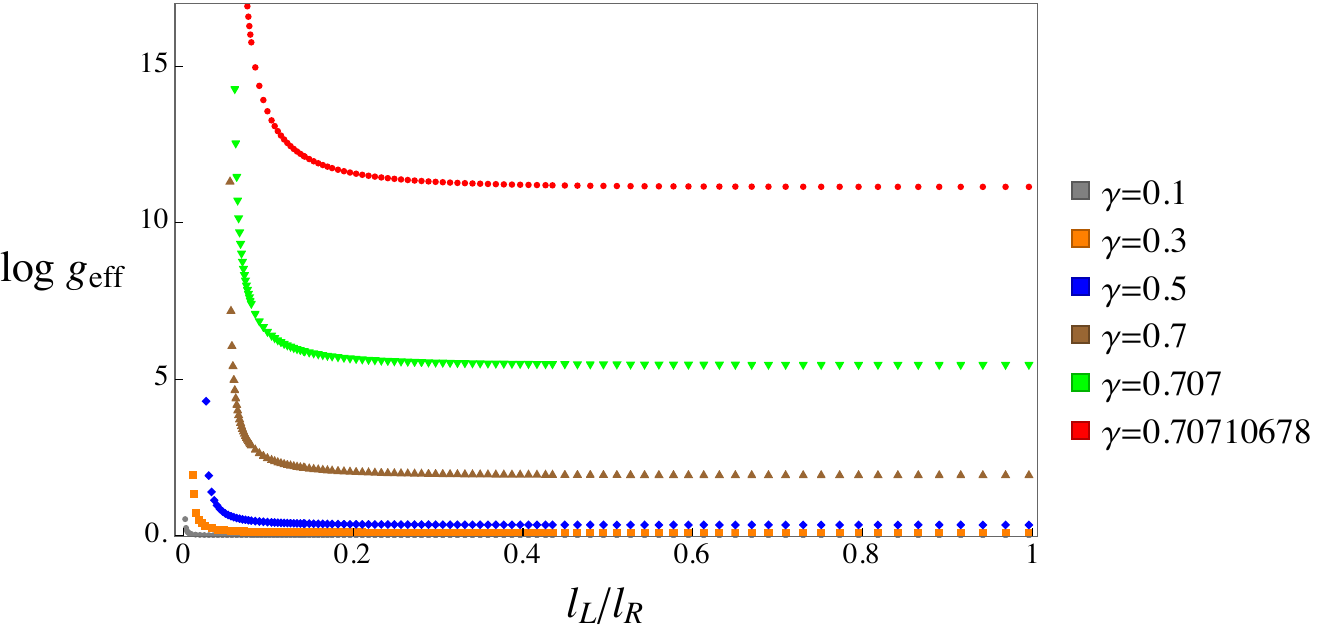}
\\[3ex]
\includegraphics[scale=0.61]{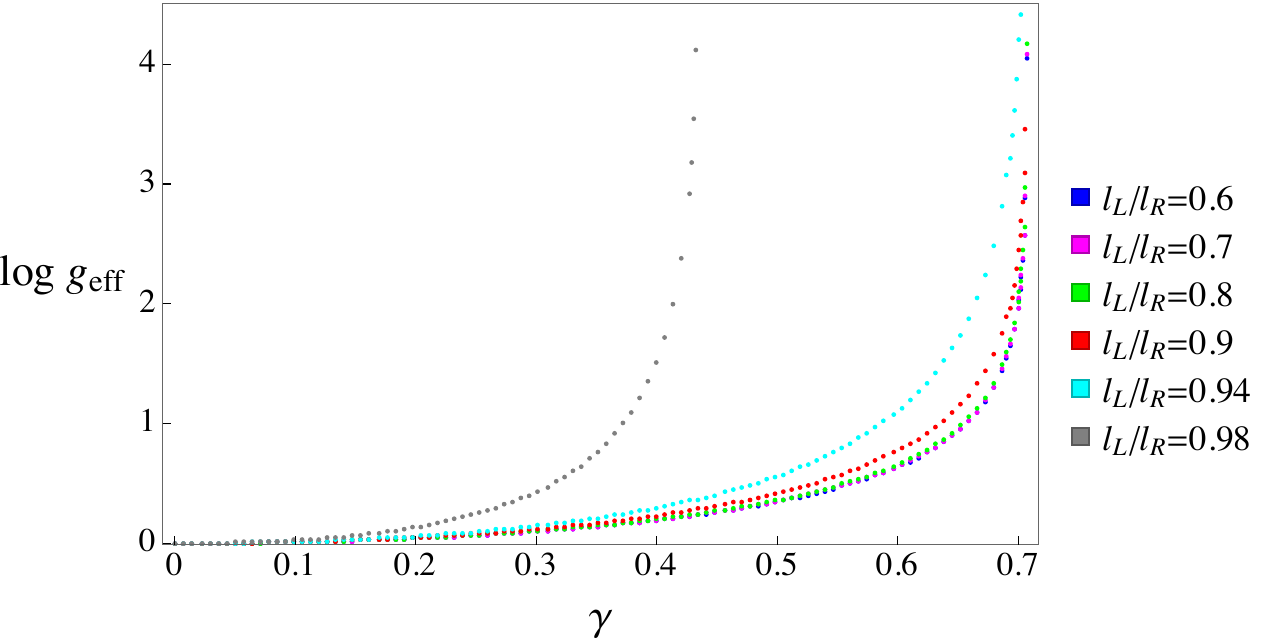}
\caption{Left: Effective interface entropy as a function of the ratio $l_L/l_R$, computed from ${\cal A}(\gamma)-{\cal A}(0)$ for various $\gamma$'s [${\cal A}(\gamma)$ is the regulated area evaluated at $\gamma$], similar to the procedure in \cite{Azeyanagi:2007qj}. As always, we have assumed $l_L\leq l_R$ so the ratio takes value from $0$ to $1$. The curves blow up on the left side near the limit of the maximally asymmetric interval, signaling a change of the $l$-dependent behavior as described in Sections \ref{sec:intro} and \ref{sec:general}, and also summarized in Figure \ref{fig:setup}. The saturation of the interface entropy on the right side of the curves is consistent with know results \cite{Azeyanagi:2007qj}; up to four significant figures, those values are $0.0101$, $0.09922$, $0.3466$, $1.956$, $5.439$, and $11.14$, from bottom up. Right: Interface entropy as a function of $\gamma$ for various $l_L/l_R$. Near $\gamma=1/\sqrt{2}$, all curves blow up but never intersect each other.}
\label{janus}
\end{figure}

For $0< c_s< \sqrt{(1+\xi)/2}$, we compute the regulated area ${\cal A}$ by numerically integrating the on-shell Lagrangian \eqref{onshelll}. The leading behavior is still $2\log l/\epsilon$ as shown in \eqref{eq:universal_log}. However, the subleading interface entropy $\log g_{\text{eff}}$ is non-universal and is plotted in Figure \ref{janus} with respect to $\gamma$ and $l_L/l_R$ (the latter is effectively controlled by $c_s$). 

Next we turn to the special case of the symmetric interval, case 1a), with $l_L=l_R$ and $c_s=0$. In this case the on-shell Lagrangian $\mathcal{L}=1$. Using the integration limits in \eqref{eq:janus_cutoff}, we find the regulated area to be simply
\begin{equation}
{\cal A}_{c_s=0} = 2\log l/\epsilon - \log \xi.
\end{equation}
The interface entropy can be viewed as boundary entropy using the folding trick. The latter has been calculated in \cite{Azeyanagi:2007qj}, and our results are consistent with theirs. Since $\xi$ takes value from $0$ to $1$, $-\log\xi>0$, namely, the interface entropy is positive.

Finally we comment on the maximally asymmetric interval, case 1b) with $l_L=0$. In this case $c_s$ takes its maximal value $\sqrt{(1+\xi)/2}.$ One can solve $x(r)$ to be
\beq
\frac{x}{x_0}=\exp\left[-\frac{1}{\sqrt{\xi}}~\text{arctanh} \left(\sqrt{\frac{1+\xi}{1+\xi\cosh 2r}}\cosh r \right)-\frac{i\pi}{2\sqrt{\xi}}\right].
\label{eq:janus_asym_x}
\eeq
From this, one can take $r\rightarrow \infty$ and obtain\footnote{One can also compute $l_L$ from the expression above, but in case 1b) one has to make $r$ approach zero from the positive side. This gives $l_L=0$ as expected.}
\begin{equation}
l_R=x_0 \exp\left(
-\frac{1}{\sqrt{\xi}}~\text{arctanh}\sqrt{\frac{1+\xi}{2\xi}}~\right).
\label{eq:janus_asym_l}
\end{equation}
We are interested in the prefactor $\sigma_1$ in the leading contribution to the entanglement entropy. This can be obtained from \eqref{deltaa2}, which we copy here for convenience:
\beq 
\frac{\delta {\cal A}}{\delta l}  = \left . {\cal L} \right |_{r=r_c^+} \frac{\delta r_c^+}{\delta l }   - \left . {\cal L} \right |_{r=r_c^0} \frac{\delta r_c^0}{\delta l}. \eeq
The first term on the right hand side will still give the usual $1/l$ factor, while the second term requires the knowledge of $\delta r_c^0/\delta l$ at the $r\rightarrow 0$ cutoff. Unlike the arguments that led to equation \eqref{sigma1}, in the special case of Janus we are able to directly solve the form of $r_c^0$. This is accomplished as follows. From \eqref{trivialcut} and \eqref{eq:AR}, we know that at the cutoff, $e^{A_*}\epsilon\approx x$. In the Janus solution, $e^{A_*}$ is simply $\sqrt{(1+\xi)/2}$. Plugging \eqref{eq:janus_asym_x} into the equation above, one finds that $r_c^0$ is related to $x_0$ in the following way:
\begin{equation}
\sqrt{\frac{1+\xi}{1+\xi\cosh 2r_c^0}} \cosh r_c^0 = -\tanh \left[ \sqrt{\xi} \log \left(e^{i\pi/2\sqrt{\xi}} \sqrt{\frac{1+\xi}{2}}\frac{\epsilon}{x_0}\right)\right].
\end{equation}
Now using \eqref{eq:janus_asym_l}, one can rewrite the above as a equation relating $l_R$ and $r_c^0$. Combining with a small-$r_c^0$ expansion on the left hand side and a small-$\epsilon$ expansion on the right hand side, one arrives at 
\begin{equation}
\left(\frac{1}{2}-\frac{\xi}{1+\xi}\right)(r_c^0)^2 = 2\left(\frac{\epsilon}{l}\right)^{2\sqrt{\xi}}.
\label{eq:janus_asym_delta}
\end{equation}
From this, one can easily find the derivative $\delta r_c^0/\delta l$. On the other hand, the on-shell Lagrangian at $r_c^0$ is
\begin{equation}
\mathcal{L}|_{r=r_c^0}=\sqrt{\frac{1+\xi\cosh 2r}{\xi (\cosh 2r-1)}}\approx \sqrt{\frac{1+\xi}{2\xi}}\frac{1}{r_c^0},
\label{eq:janus_asym_lagrangian}
\end{equation}
where we have again taken the small-$r_c^0$ expansion. Combining the two expressions \eqref{eq:janus_asym_delta} \eqref{eq:janus_asym_lagrangian} above, one finds
\begin{equation}
-\mathcal{L}|_{r=r_c^0}\frac{\delta r_c^0}{\delta l}=\sqrt{\frac{1+\xi}{2}} \frac{1}{l}= \frac{e^{A_*}}{l},
\end{equation}
leading again to $\sigma_1=\frac{c}{6}(1+e^{A_*})$ as expected from \eqref{sigma1}, where the knowledge of $r_c^0$ was unavailable.

\section{Unequal central charges}
\label{unequal}

Like in much previous work on ICFTs, we have so far focused on the case where the central charge of the CFT does not change across the interface. It is given by the same $c$ on both sides, where $c$ appeared in many of our formulas. Of course this needs not be the case. The symmetries of the ICFT allow even the central charge to jump across the interface.
This scenario has for example recently been studied in \cite{Simidzija:2020ukv,Bachas:2021fqo}. It is straightforward to adapt our techniques and results also to this case.

The central charge entered our calculations via the Brown-Henneaux relation \eqref{brh}. While we set the curvature radius of our 3d gravitational solution to $L=1$ throughout, it is important to restore this quantity in order to address what happens when the two central charges on the left and the right are not equal. With $L$ restored, the full Brown-Henneaux relation reads
\beq \label{brhf} \frac{G}{L} = \frac{3}{2c}\eeq
and so unequal central charges can be accommodated by simply allowing for asymptotically AdS$_3$ spacetimes with different curvature radii $L_L$ and $L_R$ on the two sides of the interface. In order to see how our formulas change in this case, we simply need to restore factors of $L$.

The metric of AdS$_3$ itself is still given by \eqref{poincare} with a simple overall prefactor of $L^2$. Asymptotically our holographic ICFT spacetime has to approach this form, but we can allow different curvature radii on the two sides. In terms of the pure AdS$_3$ warpfactor this means that \eqref{emptywarp} really reads
\beq e^A = L \cosh \frac{r}{L} \eeq
What is fortunate about this is that in all formulas written in terms of $e^A$, absolutely nothing changes; all the factors of $L$ are contained in $e^A$ and only appear explicitly when we plug in the asymptotic form of $e^A$ in the cutoff procedure. It is no problem to use different curvature radii for the left- and right-hand sides in this final step.

One thing to keep in mind is that now $e^{A_*}$ is no longer a number, but has unit of length. Still, all formulas in terms of $e^{A_*}$ are as before. $e^{A_*}$ always gets divided by $G$'s to yield dimensionless entanglement entropies.

It is easy to see that the area for the generic case 1) is still ${\cal A} = r_c^+ + r_c^-$, but $r_c^{\pm}$ are given by the respective curvature radius times what they were before. If we have different curvature radii $L_L$ and $L_R$, we simply get
\beq S = \frac{c_L + c_R}{6} \log(l/\epsilon) + \log g \eeq
This result is consistent with our general idea (discussed near Fig. \ref{fig:setup}) that the $\log l$ terms separately arise from the left and right boundaries of the interval. The subleading term $\log g$ can depend on $L_L$ and $L_R$ as well.

Case 1b) now splits into two different scenarios; the one-sided interval can either be to the ``right" or the ``left" of the interface. For concreteness, let us chose the case where it is to the right. To get the formulae for the interval to the left of the interface, one simply replaces $c_R$ with $c_L$. Case 2) is still uniquely defined. With unequal tensions, the holographic scenario now yields
\beq 
\sigma_1 = \frac{c_R}{6} + f, \quad \sigma_2 = f
\label{newupshot}
\eeq
where
\beq
f = \frac{e^{A_*}}{4G} = \frac{c_L}{6} \frac{e^{A_*}}{L_L} = \frac{c_R}{6} \frac{e^{A_*}}{L_L}.
\eeq
Clearly these formulas reduce to \eqref{upshot} from before once we equate $L_L=L_R$ and hence $c_L=c_R$. In short, the ``natural" way of writing the interface contribution never had a central charge in there. The interface contribution $f$ is given by the curvature radius of the minimal slice in Planck units. In the case we had a single curvature radius, we were able to write $e^{A_*}$ as a pure number times $L$, so it was natural to write $f$ as $c$ times a number. But for general $c_L$ and $c_R$, it is more natural to just assign an $f$ to the defect as above and to not even bother relating it to the central charges. Note that $c_L$ and $c_R$ are intrinsic properties of the interface, so there is one number $f$, the contribution of the defect. It does not get ``weighted'' by $c$.

Once again, the general results for cases 2) and 1b) are still consistent with our result that the $\log l$ terms are dominated by the interval endpoints.

\section*{Acknowledgments}
We would like to thank Costas Bachas, Saba Baig, Ilka Brunner and Michael Gutperle for very helpful email exchanges. Zhu-Xi appreciates conversations with Wenjie Ji on the entanglement entropy for multiple disjoint intervals. We thank the JHEP referee for drawing our attention to the case with unequal central charges.
This work was supported, in part, by  a grant from the Simons Foundation (Grant 651440, AK and LB, the Simons Collaboration on Ultra-Quantum Matter). 

\begin{appendix}
    \section{Alternative  Derivation of \texorpdfstring{$S_{EE}$}{} in the RS braneworld}
\label{app}

In this appendix we present an alternative way to study the EEs in RS braneworlds, following the methods employed in recent studies of entanglement islands in RS braneworlds, in particular\footnote{In \cite{Geng:2020fxl}, our case 2) inter-CFT entanglement entropy was referred to as ``left-right entanglement". Here we refrain from using this term in order to avoid confusion with the left-right entanglement studied for example in \cite{PandoZayas:2014wsa,Lundgren:2014qua, Das:2015oha}, which is a different quantity.} \cite{Chen:2020uac,Geng:2020fxl}. For this purpose, we first find the most general form of the RT surface away from the brane and then impose an non-trivial gluing condition across the brane. For this purpose, it is most convenient to work in the Poincar\'e patch metric \eqref{poincare}, as this makes the RT surface particularly easy (every RT surface in AdS$_3$ is a semicircle centered on the asymptotic boundary in these coordinates).

Using the change of coordinates \eqref{coc}, a brane located at a constant $r_*$ maps to a brane located at
\beq z = - \tan(\theta) y \eeq
with
\beq \tanh^{-1}(\cos \theta ) = r_*  .\eeq
This time we are looking for a surface $t=0$ and $y(z)$ with Lagrangian
\beq
{\cal L} = \frac{1}{z} \sqrt{1 + (y')^2}. \eeq
As in \cite{Azeyanagi:2007qj}, where the same system was studied in the $\rho$-$\tau$ coordinates, the equations of motion can be integrated to
\beq y' = \pm \frac{z}{\sqrt{a^2-z^2}}  .\eeq
and hence
\beq
\label{explicitsolution}
y = \pm \sqrt{a^2-z^2} + y_0.
\eeq
As advertised, the RT surfaces are semicircles.

We will use coordinates where on both sides of the brane the spacetime metric is just given by \eqref{poincare} with $y$ increasing away from the brane.\footnote{$y$-coordinates are denoted by $y_R$ and $y_L$ on the left and right sides of the brane, respectively. As we will see later, $y_R,y_L\in [\min\{0,y_*\},\infty)$.} This is a very discontinuous choice of coordinates. It simplifies the RT surfaces (so we do not have to keep track of extra signs in $y$), but as we will see it slightly complicates the junction conditions.

Near $z=0$ the RT surface approaches the boundary at $l_{L/R}$ respectively. This tells us
\beq l_{L/R} = y_{0,L/R} + a_{L/R} .\eeq

The RT surface intersects the brane at a point $(y_*,z_*)$ with $z_* = - \tan \theta y_*$. Clearly the RT surface has to be continuous, so both sides have to yield the same $y_*$. To understand the jump equation on the derivative, it helps to write the Lagrangian for $y(\xi)$, $z(\xi)$ with independent worldvolume parameter $\xi$ of fixed coordinate length:
\beq {\cal L} = \frac{1}{z} \sqrt{\dot{y}^2 + \dot{z}^2} \eeq
where dots are $\xi$ derivatives. When deriving the equations of motion, we are left with a boundary term
\beq \delta S = (eom) + \left .  \frac{1}{z} \frac{\dot{y} \delta y_* + \dot{z} \delta z_*}{\sqrt{\dot{y}^2 + \dot{z}^2}}  \right |_{brane} . \eeq
Dots signify $\xi$ derivatives.
Since we get a contribution like this from both sides (where again, the fact that $y$ is chosen to increase away from the defect in both directions means we {\it add} the contributions from both sides with the same sign), the boundary condition reads
\beq \label{bc2} \left  ( \frac{\dot{y}_L}{\sqrt{\dot{y}_L^2 + \dot{z}_L^2}} + \frac{\dot{y}_R}{\sqrt{\dot{y}_R^2 + \dot{z}_R^2}} \right ) \delta y_* + \left ( \frac{\dot{z}_L}{\sqrt{\dot{y}_L^2 + \dot{z}_L^2}} - \frac{\dot{z}_R}{\sqrt{\dot{y}_L^2 + \dot{z}_L^2}} \right ) \delta z_* = 0 \eeq
Again, the peculiar sign in the $\delta y_*$ term comes from our unusual choice of coordinate in this direction.
For the endpoint to be located on the brane, we need $\delta z_* = -(\tan \theta) \delta y_*$. Futhermore, now that we derived the boundary conditions we can again go to a parameterization where $\xi=z$ and so $\dot{z}=1$ and $\dot{y}=y'$, so that \eqref{bc2} becomes simply
\beq y_L' = y_R'. \label{boundarycondition} \eeq
In these coordinates, the RT surface is continuous and so is its derivative.

It is tedious but straightforward to verify that our formal solution \eqref{rssolution}, when translated to $y$-$z$ coordinates, indeed gives semi-circles. While this can be done analytically, the resulting expressions of $y(z)$ are very tedious to write down. The boundary condition \eqref{boundarycondition} has to be automatically obeyed in the piecewise-defined solution \eqref{rssolutionintegrated}. 

Now we figure out $y_L$, corresponding to $r<0$. Denoting the expression $\frac{c_s\sinh r_*}{\sqrt{\cosh^2r_*-c_s^2}}$ in the exponent in \eqref{x0relation} as $\beta$ and the expression $\frac{c_s\sinh(r+r_*)}{\sqrt{\cosh^2(r+r_*)-c_s^2}}=-\frac{y_L}{z}\frac{c_s}{\sqrt{y_L^2/z^2+1-c_s^2}}$ in the exponent in the second row of \eqref{rssolutionintegrated} as $\tilde{\beta}_+$, we get the translated equations
\begin{equation}
    \left(\frac{z\cosh(r+r_*)}{x_0}\right)^2=\frac{1+\tilde{\beta}_+}{1-\tilde{\beta}_+}\left(\frac{1-\beta}{1+\beta}\right)^2,
\end{equation}
but since $l_Rl_L=x_0\tilde{x}_0$ from \eqref{newalpharelation}, the equation above becomes
\begin{equation}
y_L^2+z^2=l_Rl_L\frac{1+\tilde{\beta}_+}{1-\tilde{\beta}_+}\frac{1-\beta}{1+\beta}:=K\frac{1+\tilde{\beta}_+}{1-\tilde{\beta}_+}.
\end{equation}
Two out of four solutions are
\begin{equation}
    y_L=-\sqrt{\frac{K \left(c_s^2+1\right)\pm2 \sqrt{K c_s^2 \left(K+\left(c_s^2-1\right) z^2\right)}}{1-c_s^2}-z^2}.
\end{equation}
Now by setting $C:=K-z^2(1-c_s^2),\,D:=c_s\sqrt{K}$ and noticing that $\sqrt{C}<D$, we get a perfect square $\big(\sqrt{C}\pm D\big)^2$ under the outermost square root, and
\begin{equation}
\label{eq:yL}
    y_L=-\frac{c_s\sqrt{K}}{\sqrt{1-c_s^2}}\mp\sqrt{\frac{K}{1-c_s^2}-z^2}.
\end{equation}
The $\pm$
sign accounts for the fact that a solution in this parameterization has two branches, so the full solution is the union of both. 
These are indeed parts of semicircles.

To get $y_R$, corresponding to $r>0$, we denote the expression $\frac{c_s\sinh(r-r_*)}{\sqrt{\cosh^2(r-r_*)s-c_s^2}}$ in the exponent in the first row of \eqref{rssolutionintegrated} by $\tilde{\beta}_-$, and go through an easier exercise. We find out $y_R$ to be the expression in \eqref{eq:yL} with $K=1$, and now the boundary condition \eqref{boundarycondition} is trivially satisfied. 

\end{appendix}

\bibliographystyle{JHEP}
\bibliography{interfaces.bib}

\end{document}